\documentclass[letterpaper,showpacs,preprintnumbers,superscriptaddress,aps,nofootinbib,perprint,10pt]{revtex4}
\usepackage{amssymb}\usepackage[centertags]{amsmath}\usepackage{txfonts}\usepackage{epsfig}\usepackage{bm}
\usepackage{color}\usepackage{graphicx,graphics}\usepackage{multirow}\usepackage{float}\usepackage{ulem}
\usepackage[dvipdfm,pdfstartview=FitH]{hyperref}\usepackage{setspace}\usepackage{slashed}\usepackage{booktabs}
\hypersetup{colorlinks=true, citecolor=blue, linkcolor=red, filecolor=black,urlcolor=blue}
\allowdisplaybreaks[4]

\begin{document}

\title{Parity-odd Parton Distribution Functions from $\theta$-Vacuum}

\author{Weihua Yang } 

\affiliation{National key Laboratory of Science and Technology on Reliability and Environmental Engineering, Beijing Institute of Spacecraft Environment Engineering, Beijing 100094, China}

\begin{abstract}
    Quantum chromodynamics is a fundamental non-abelian gauge theory of strong interactions. The physical quantum chromodynamics vacuum state is a linear superposition of the $n$-vacua states with different topological numbers. Because of the configuration of the gauge fields, the tunneling events can induce the local parity-odd domains. Those interactions that occur in these domains can be affected by these effects. Considering the hadron (nucleon) system, we introduce the parity-odd parton distribution functions in order to describe the parity-odd structures inside the hadron in this paper. We obtain 8 parity-odd parton distribution functions at leading twist for spin-1/2 hadrons and present their properties. By introducing the parity-odd quark-quark correlator, we find the parity-odd effects vanish from the macroscopic point of view. In this paper, we consider the high energy semi-inclusive deeply inelastic scattering process to investigate parity-odd effects by calculating the spin asymmetries.
\end{abstract}


\maketitle

\section{Introduction}\label{S:introduction}

Quantum chromodynamics (QCD) is a fundamental non-abelian gauge theory of strong interactions whose Lagrangian deals with quarks, gauge fields (gluons) and their interactions. QCD has two outstanding features, asymptotic freedom and quark confinement. Thanks to the asymptotic freedom, many high energy reactions can be calculated in the form of factorization theorems \cite{Collins:1989gx}, which separate the cross sections into calculable hard parts and non-perturbative soft parts. The soft parts are often factorized as parton distribution functions (PDFs) and fragmentation functions (FFs) according to the specific reaction processes. When three dimensional, i.e., the transverse momentum dependent (TMD) PDFs and FFs are considered, the sensitive quantities studied in experiments are different azimuthal asymmetries. In this paper, we only consider PDFs. These (TMD)PDFs are defined via the quark-quark correlator which satisfies the constraints imposed by hermiticity, time reversal and parity conservation. However, the parity conservation can be violated locally by the parity-odd tunneling events induced by the non-trivial configurations of the gauge fields in the local parity-odd domains.

It is known that the true physical QCD vacuum state is a linear superposition of the $n$-vacua states with different topological numbers \cite{tHooft:1976rip,Jackiw:1976pf,Callan:1976je}. The vacuum transition amplitudes is just determined by the topological charge (density) which is equal to the $\theta$-term, $\theta \tilde G^{\mu\nu}G_{\mu\nu}$. By applying the path integral formation, one finds the $\theta$-term can be added to the QCD Lagrangian \cite{Fujikawa:1979ay,Peskin:1995ev,Crewther:1978zz} and results in the strong CP problem and the axial vector current anomaly \cite{Peccei:2006as,Cheng:1987gp,Adler:1969gk}.
Though QCD is parity violated, the measurements of electric dipole moment of neutron indicate that the parity violation is $local$~\cite{Baker:2006ts}. Therefore the perturbative QCD still respects to the global parity conservation constraint. Parity-odd effects induced by the random fluctuations in the parity-odd domains could be observed via multi-particle correlations, handedness correlation and azimuthal asymmetries~\cite{Kharzeev:1998kz,Efremov:1995ff,Kang:2010qx,Yang:2019rrn}.
Beyond that, in non-central heavy ion collisions the chiral magnetic effect indicates that tunneling event can yield charge-separation when parity-odd domain interacts with the very large magnetic field \cite{Kharzeev:2004ey,Adamczyk:2013hsi}.
Even in neutron the $\theta$-term can generate charge-separation which gives rise to the neutron electric dipole moment (nEDM) \cite{Faccioli:2004ys}.

When tunneling events induced by non-trivial configurations of the gauge fields are taken into consideration, parity-odd PDFs can emerge.
Recently, we have presented our discussions of parity-odd FFs \cite{Yang:2019gdr}. As an extension, we focus our attentions on the PDFs in this paper. We obtain 8 parity-odd PDFs at leading twist for spin-1/2 hadrons by decomposing the quark-quark correlator. We also present positivity bounds of these one dimensional PDFs. Following a brief introduction to the non-trivial $\theta$-vacuum which induces parity-odd PDFs in Sect. \ref{S:thetavacuum}, we present the decomposition of the quark-quark correlator at leading twist in Sect. \ref{S:PDFS}. We obtain 8 parity-odd PDFs at leading twist for spin-1/2 hadrons. In Sect. \ref{S:Properties} we present the positivity bounds of these parity-odd PDFs and introduce the operator definition of the parity-odd correlator. In Sect. \ref{S:Applications}, we calculate the spin asymmetries in semi-inclusive deeply inelastic scattering (SIDIS) process. Finally, a brief summary is given in Sect \ref{S:summary}.

\section{The $\theta$-term in QCD Lagrangian}\label{S:thetavacuum}

The axial vector current is not conserved in QCD, it can obtain the quantum corrections from the triangle diagrams \cite{Adler:1969gk}. The divergence of the axial vector current is given by
\begin{align}
  \partial_\mu j^{\mu5}=\frac{g^2}{64\pi^2}\varepsilon^{\alpha\beta\mu\nu}G_{\alpha\beta}G_{\mu\nu}, \label{f:divergenceresult}
\end{align}
where $g$ is the strong interaction coupling constant. $G_{\alpha_\beta}$ is the full field strength tensor of the gauge field. The pseudoscalar term can be written as a total divergence
  $\varepsilon^{\alpha\beta\mu\nu}G_{\alpha\beta}G_{\mu\nu}=2\partial_\mu K^\mu$, 
with
  $K^\mu = \varepsilon^{\mu\alpha\beta\gamma}A_\alpha^a \big[G_{\beta\gamma}^a-\frac{g}{3}f^{abc}A_\beta^b A_\gamma^c\big]$. 
Using the boundary condition $A_\mu=0$ at spatial infinity, one finds that the axial vector current satisfies the following equation,
\begin{align}
  \int d^4x \partial j^{\mu5}= \frac{g^2}{32\pi^2} \int d^4x \partial_\mu K^\mu =\frac{g^2}{32\pi^2} \int d\sigma_\mu K^\mu =0. \label{f:axialequation}
\end{align}
However, this equation is not correct because the boundary condition is not being chosen correctly. 't Hooft argued that $A_\mu$ should be a pure gauge at spatial infinity \cite{tHooft:1976rip}.

It is known that under a gauge transformation $\Omega$ the gauge field transforms as
\begin{align}
  A_\mu\to \Omega A_\mu \Omega^{-1} + \frac{i}{g} (\partial_\mu \Omega) \Omega^{-1}. \label{f:Atrans}
\end{align}
Putting $A_\mu=0$ into this equation yield the vacuum configuration of pure gauge,
  $A_\mu^{pure} =\frac{i}{g} (\partial_\mu \Omega) \Omega^{-1}$. 
In the temporal gauge $A_0 =0$, one can classify these vacuum configuration by requiring $\Omega$ going to unity as $r\to \infty$,
\begin{align}
  \Omega\to e^{i2\pi n}, \quad r\to \infty,  \quad   n= 0, \pm1, \pm 2, \cdots. \label{f:omega}
\end{align}
This suggests that the surface integral over the current $K^\mu$ in Eq. ({\ref{f:axialequation}}) does not vanish. The integer $n$ which is known as winding number is determined by an integral over the pure gauge fields,
\begin{align}
  n=\frac{g^2}{32\pi^2}\int d^3r K^0_{n}, \quad  K^0_{n} =-\frac{g}{3}f_{ijk}\varepsilon_{abc}A^{ia}_{n}A^{jb}_{n}A^{kc}_{n}. \label{f:nwinding}
\end{align}

A transition occurs from a configuration with $n_-$ at $t=-\infty$ to one with $n_+$ at $t=+\infty$ can be  expressed by,
\begin{align}
  \nu=n_+-n_- = \frac{g^2}{32\pi^2} \int d\sigma_\mu K^\mu\Big|^{t=+\infty}_{t=-\infty} =\frac{g^2}{32\pi^2} \int d^4x \tilde G^{\mu\nu}G_{\mu\nu}, \label{f:vwinding}
\end{align}
where $\nu$ is the difference of the winding numbers, $\tilde G^{\mu\nu}=\frac{1}{2}\varepsilon^{\alpha\beta\mu\nu}G_{\alpha\beta}$. It is known the true vacuum is a linear superposition of the $n$-vacua,
  $|\theta \rangle =\sum_n e^{-in\theta} |n \rangle$, 
with $\theta$ being a real number. A transition between two vacua at $t=- \infty$ and $t=+ \infty$,
\begin{align}
  \langle \theta (+\infty)| \theta(-\infty)\rangle =\sum_\nu e^{i\nu\theta} \sum_n \langle (n+\nu)(+\infty)| n(-\infty)\rangle,  \label{f:thetatrans}
\end{align}
can be expressed through the path integral formation:
\begin{align}
  \langle \theta (+\infty)| \theta(-\infty)\rangle =\sum_\nu \int \delta A e^{iS_{eff}[A]}\delta\left(\nu-\frac{g^2}{32\pi^2}\int d^4x \tilde G^{\mu\nu}G_{\mu\nu}\right), \label{f:thetapath}
\end{align}
where $\frac{g^2\theta}{16\pi^2}\tilde G^{\mu\nu}G_{\mu\nu}$ has been added to the customary QCD Lagrangian:
\begin{align}
  \mathcal{L}=-\frac{1}{4}(G_{\mu\nu})^2+\bar\psi(i\slashed D-m)\psi + \frac{\theta g^2}{32\pi^2} \tilde G^{\mu\nu}G_{\mu\nu}. \label{f:thetalagrangian}
\end{align}
The covariant derivative $D_\mu = \partial_\mu -igA_\mu$. By utilizing Eq. (\ref{f:divergenceresult}) with $j^{\mu5}=\bar \psi\gamma^\mu\gamma^5\psi$, the Lagrangian can be rewritten as:
\begin{align}
  \mathcal{L} &=-\frac{1}{4}(G_{\mu\nu})^2+\bar\psi(i\slashed \partial-m)\psi + \bar\psi\gamma^\mu(gA_\mu - \tilde\theta_\mu\gamma^5)\psi , \label{f:Ldelta}
\end{align}
where $\tilde\theta_\mu =\partial_\mu \theta $. Here we assume that $\theta=\theta(x)$ is non-zero. Since $\theta$ is a pseudoscalar field, $\tilde\theta$ can be taken as a pseudovector. $\tilde \theta$ is different from the vector field potential, $A_\mu$, it plays a role of the potential coupling to the axial vector current which is determined by the topological charge (density) Q ($\partial_\mu j^{\mu5}=Q$).

It is can be seen that the $\theta$-term in the QCD Lagrangian is parity violated ($\tilde G^{\mu\nu}G_{\mu\nu}=-4\vec B\cdot \vec E$). It can form domains in which interactions are affected by the $\theta$-vacuum tunneling events. A study shows that the $\theta$-term can generates an effective repulsion between quarks and antiquarks in these parity-odd domains \cite{Faccioli:2004ys}. As a consequence, the repulsion gives rise to the EDM. Considering the quark distributions, the $\theta$-vacuum tunneling events can also give rise to the parity-odd PDFs. In the following context we introduce these parity-odd PDFs for spin-1/2 hadrons and shown some properties of them.


\section{Parton distribution functions}\label{S:PDFS}


In the quantum field theory, the PDFs are defined via the quark-quark correlator which is given by
\begin{align}
  \hat\Phi(k,p,S)= \frac{1}{2\pi}\int d^4\xi e^{ik\xi} \langle N,S|\bar \psi(0)\mathcal{L}(0,\xi)\psi(\xi) |N,S \rangle, \label{f:correlatorE}
\end{align}
where $k$ denotes the quark momentum, $p$ and $S$ are the momentum and spin of the hadron. $\mathcal{L}(0,\xi)$ is the gauge link which keeps the correlation function gauge invariant. However, the appearance of gauge links has no influence on the following discussions. Therefore, we just omit it for simplicity in the following context.
The correlator $\hat\Phi(k,p,S)$ defined in Eq. (\ref{f:correlatorE}) depends on the quark 4-momentum $k$. We know that the TMD PDFs are defined via the three dimensional quark-quark correlator. To obtain this, we first introduce the light-cone unit vectors $n=(0,1,\vec 0_T), \bar n=(1,0,\vec 0_T)$ which satisfy $n\cdot \bar n =1, n^2=\bar n^2=0$. We choose the hadron's momentum as the z-direction. In this case, the polarization vector, the momenta of the hadron and the quark can be parametrized as
\begin{align}
  &S^\mu = \lambda_h \frac{p^+}{M}\bar n^\mu -\lambda_h \frac{M}{2p^+} n^\mu + S_T^\mu, \label{f:S}\\
  &p^\mu= p^+\bar n^\mu +\frac{M^2}{2p^+} n^\mu, \label{f:p} \\
  &k^\mu= k^+\bar n^\mu +\frac{k^2+\vec k_T^2}{2k^+} n^\mu + k^\mu_T, \label{f:k}
\end{align}
where $k^+=xp^+$, $\lambda_h$ is the helicity of the hadron.
By integrating over $k^-=\frac{k^2+\vec k_T^2}{2k^+}$ in the light-cone coordinates, we can obtain the TMD correlator which is given by
\begin{align}
  \hat\Phi(k_T, p, S)&=\frac{1}{2\pi}\int d\xi^-d^2\xi_T e^{ixp^+\xi^- - i\vec k_T \vec \xi_T} \langle N,S|\bar \psi(0,0,\vec 0_T) \psi(0,\xi^-, \vec \xi_T)|N,S \rangle .\label{f:TMDcorrelator}
\end{align}
We note that the correlator $\hat\Phi(k_T, p, S)$ is a $4\times 4$ matrix in Dirac space depending on the hadron state and can be decomposed in terms of the Dirac matrices, i.e., $\Gamma = \{I, i\gamma^5, \gamma^\alpha, \gamma^5\gamma^\alpha, i\sigma^{\alpha\beta}\gamma^5$\}, and the corresponding coefficient functions. We decompose the correlator in the following way,
\begin{align}
  \hat\Phi&=I\Phi+i\gamma^5\tilde\Phi +\gamma^\alpha \Phi_\alpha +\gamma^5\gamma^\alpha \tilde\Phi_\alpha+i\sigma^{\alpha\beta}\gamma^5\Phi_{\alpha\beta} \nonumber\\
  &+ i\gamma^5\Phi^p+\tilde\Phi^p +\gamma^5\gamma^\alpha\Phi_\alpha^p +\gamma^\alpha \tilde\Phi_\alpha^p+i\sigma^{\alpha\beta}\Phi_{\alpha\beta}^p, \label{f:PhiD}
\end{align}
where we have omitted the argument $(k_T, p, S)$, superscript $p$ is used to represent the extra functions except for the normal ones.
We can further expand these coefficient functions according to their Lorentz transformation properties in terms of the basic Lorentz covariants constructed from basic variables at hand. The coefficient functions are expressed as the sum of the basic Lorentz covariants multiplied by (pseudo)scalar functions which are known as the TMD PDFs. From Eqs. (\ref{f:S})-(\ref{f:k}), we see only $1, M, \lambda_h, \bar n_\alpha, k_{T\alpha}, S_{T\alpha}, g_{T\alpha\beta}$ and $\varepsilon_{T\alpha\beta}$ can be used to construct Lorentz covariants at leading twist. To be explicit, we list all the covariants in Table. \ref{T:Lorentz}.

\begin{table}[t]
\begin{center}
\renewcommand\arraystretch{1.5}
\newcommand{\tabincell}[2]{\begin{tabular}{@{}#1@{}}#2\end{tabular}}
\begin{tabular}{|c||c|c|}
  \hline
   & P-even (PDF) & P-odd (PDF)  \\ \hline \hline
  vector & $\bar n_\alpha,\quad \bar n_\alpha\varepsilon_{TkS}$ & $\gamma^5\bar n_\alpha,\quad \gamma^5 \bar n_\alpha\varepsilon_{TkS}$ \\  \hline
  pseudo-vector & ~~ $\lambda_h\bar n_\alpha,\quad (k_T\cdot S_T)\bar n_\alpha$ ~~ & ~~ $\gamma^5\bar n_\alpha \lambda_h,\quad \gamma^5\bar n_\alpha(k_T\cdot S_T)$ ~~  \\  \hline
  ~~pseudo-tensor~~ & ~~ \tabincell{c}{$\bar n_\alpha \varepsilon_{Tk\beta},\quad \bar n_\alpha S_{T\beta}$,\\ $ \lambda_h\bar n_\alpha k_{T\beta},\quad (k_T\cdot S_T)\bar n_\alpha k_{T\beta}$} ~~ & ~~ \tabincell{c}{$\bar n_\alpha k_{T\beta},\quad \bar n_\alpha \varepsilon_{TS\beta}$,\\ $\lambda_h\bar n_\alpha \varepsilon_{Tk\beta},\quad (k_T\cdot S_T)\bar n_\alpha \varepsilon_{Tk\beta}$ }~~ \\  \hline
\end{tabular} \caption{Basic Lorentz covariants used to construct these coefficient functions.}\label{T:Lorentz}
\end{center}
\end{table}

We take the pseudo-vector term for example. Using the third line in Table. \ref{T:Lorentz}, we have
\begin{align}
  \hat\Phi&=\lambda_h\gamma^5\gamma^\alpha\bar n_\alpha  A+\gamma^5\gamma^\alpha\bar n_\alpha (k_T\cdot S_T)B+ \lambda_h\gamma^5\gamma^\alpha\gamma^5\bar n_\alpha C + \gamma^5\gamma^\alpha\gamma^5\bar n_\alpha(k_T\cdot S_T)D \nonumber\\
  &=\lambda_h\gamma^\alpha\gamma^5\bar n_\alpha  A+\gamma^\alpha\gamma^5\bar n_\alpha (k_T\cdot S_T)B- \lambda_h\gamma^\alpha\bar n_\alpha C - \gamma^\alpha\bar n_\alpha(k_T\cdot S_T)D,
\end{align}
where $A, B, C, D$ are (pseudo)scalar functions. In keeping with conventions, we redefine $g_{1L}\equiv A, g_{1T}^\perp \equiv MB, v_{1L}\equiv -C$ and $v_{1T}^\perp \equiv -MD$. Thus we obtain Eqs. (\ref{f:phi5}) and (\ref{f:phi5P}). The other PDFs can be obtained in the similar way, we present them in the following.
As a consequence of parity constraint, at leading twist the coefficient functions decomposed from the quark-quark correlator can be expanded as
\begin{align}
  &\Phi_\alpha(x, k_T)=\bar n_\alpha\bigg[f_1(x, k_T)+\frac{\varepsilon_{T kS}}{M}f^\perp_{1T}(x, k_T)\bigg],\label{f:phi}\\
  &\tilde\Phi_\alpha(x, k_T)=\bar n_\alpha\bigg[\lambda_h g_{1L}(x, k_T)+\frac{k_T \cdot S_T}{M}g^\perp_{1T}(x, k_T)\bigg],\label{f:phi5}\\
  &\Phi_{\rho\alpha}(x, k_T)=\bar n_\alpha\bigg[\frac{\varepsilon_{T k\rho}}{M}h^\perp_{1}(x, k_T) +S_{T\rho}h_{1T}(x, k_T)+\frac{k_{T\rho}}{M}h^\perp_{1S}(x, k_T)\bigg].\label{f:phiodd}
\end{align}
Here we defined the shorthanded notation $h^\perp_{1S}=\lambda_h h^\perp_{1L}+\frac{k_T \cdot S_T}{M}h^\perp_{1T}$. In Eqs. (\ref{f:phi})-(\ref{f:phiodd}), we use $f, g$ and $h$ to denote the unpolarized, longitudinal polarized and transversely polarized quark distribution functions in hadrons. Subscript $1$ stands for the leading twist and $L, T$ specify the polarizations of the hadron. Superscript $\perp$ indicates that these PDFs are all dependent on the transverse momentum of the quark. By integrating over $k_T$, we can obtain the one dimensional PDFs,  $f_1(x)$, $g_{1L}(x)$ and $h_{1T}(x)$.

If the parity constraint is released, the parity-odd PDFs which are induced by the non-trivial $\theta$-vacuum tunneling events can emerge. In this case, we can obtain the parity-odd PDFs by decomposing the coefficient functions,
\begin{align}
  &\Phi^P_\alpha(x, k_T)=\bar n_\alpha\bigg[u_1(x, k_T)+\frac{\varepsilon_{T kS}}{M}u^\perp_{1T}(x, k_T)\bigg],\label{f:phiP}\\
  &\tilde\Phi^P_\alpha(x, k_T)=\bar n_\alpha\bigg[\lambda_h v_{1L}(x, k_T)+\frac{k_T \cdot S_T}{M}v^\perp_{1T}(x, k_T)\bigg],\label{f:phi5P}\\
  &\Phi^P_{\rho\alpha}(x, k_T)=\bar n_\alpha\bigg[\frac{k_{1T\rho}}{M_1}w^\perp_{1}(x, k_{1T}) +\varepsilon_{T S_1\rho}w_{1T}(x, k_{1T})+\frac{\varepsilon_{T k_1\rho}}{M_1}w^\perp_{1S}(x, k_{1T})\bigg].\label{f:phioddP}
\end{align}
where superscript $P$ is used to represent the parity-odd quantities, $w^\perp_{1S}=\lambda_h w^\perp_{1L}+\frac{k_T \cdot S_T}{M}w^\perp_{1T}$. $u, v$ and $w$ are used to denote the parity-odd unpolarized, longitudinal polarized and transversely polarized quark distribution functions in hadrons. 
They have a one-to-one correspondence to parity-even PDFs, $f, g$ and $h$. Hence they can be discussed simultaneously.
The one dimensional parity-odd PDFs are similar to the parity-even ones, they are $u_1(x)$, $v_{1L}(x)$ and $w_{1T}(x)$. Here we only present the leading twist PDFs. For the higher twist parity-even and parity-odd PDFs, they can also be obtained in a similar way \cite{Chen:2016moq}. We do not repeat the decompositions in this paper.



\section{Properties of parton distribution functions}\label{S:Properties}

By using the optical theorem, the leading twist PDFs can be expressed in terms of quark-hadron forward amplitudes, $\mathcal{A}_{\Lambda\lambda,\Lambda'\lambda'}$, where $\lambda, \lambda' (\Lambda, \Lambda')$ are the helicities of the incoming and outgoing quark (hadron) in the u-channel process. By using the time-reversal, helicity conservation and parity constraints, only three independent amplitudes survival,
\begin{align}
  &\mathcal{A}_{++,++}, &&\mathcal{A}_{+-,+-}, &\mathcal{A}_{++,--}.
\end{align}
It is worthwhile to note that the parity constraint used to eliminate the correlated amplitudes puts no constraint on the parity-odd amplitudes (vertices). As mentioned before, the parity violation in QCD is local and these parity-odd amplitudes are localized in small domains. By averaging over all the parity-odd fluctuations, all of these parity violated effects can be reduced completely at the macroscopic level. In other words, the parity-odd quantities contribute nothing to global quantities when average over all the contributions.

\begin{figure}
  \centering
  \includegraphics[width=6cm]{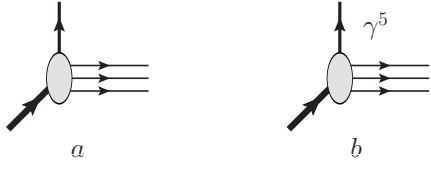}\\
  \caption{The parity-even (a) and parity-odd (b) vertices.}\label{Fig:pvertices}
\end{figure}

To describe the forward amplitudes, we define the parity-even and parity-odd quark-hadron vertices $a_{\Lambda\lambda}$ and $a^P_{\Lambda\lambda}$, see Fig.~(\ref{Fig:pvertices}). Thus, we can express the amplitudes in terms of the quark-hadron vertices,
\begin{align}
  \mathcal{A}_{++,++} &\sim \sum_X\big(a^*_{++}+a^{P*}_{++}\big)\big(a_{++}+a^{P}_{++}\big) = \sum_X\big[2a^*_{++}a_{++}+ 2 Re (a^{P*}_{++}a_{++})\big], \label{f:Azzzz} \\
  \mathcal{A}_{+-,+-} &\sim \sum_X\big(a^*_{+-}+a^{P*}_{+-}\big)\big(a_{+-}+a^{P}_{+-}\big) = \sum_X\big[2a^*_{+-}a_{+-}+ 2 Re (a^{P*}_{+-}a_{+-})\big], \label{f:Azfzf} \\
  \mathcal{A}_{++,--} &\sim \sum_X\big(a^*_{--}+a^{P*}_{--}\big)\big(a_{++}+a^{P}_{++}\big) = \sum_X\big[2a^*_{--}a_{++}+ 2 Re (a^{P*}_{--}a_{++})\big]. \label{f:Affzz}
\end{align}
We can see that these first terms on the right hand side of the second equalities in Eqs. (\ref{f:Azzzz})-(\ref{f:Affzz}) respectively are corresponding to the parity-even PDFs while the others correspond to the parity-odd ones. To obtain these relations, we have used $a^*_{\Lambda\lambda}a_{\Lambda'\lambda'}= a^{P*}_{\Lambda\lambda}a^P_{\Lambda'\lambda'}$. For the parity-odd amplitudes, we have $\mathcal{A}_{\Lambda\lambda,\Lambda'\lambda'} \sim 2 Re (a^{P*}_{\Lambda'\lambda'}a_{\Lambda\lambda})$.
By using the optical theorem to relate the amplitudes to the three leading-twist one dimensional PDFs, we can obtain
\begin{align}
  &f_{1\ }+u_{1\ } \sim Im(\mathcal{A}_{++,++}+\mathcal{A}_{+-,+-}) \sim \sum_X2\big[a^*_{++}a_{++}+a^*_{+-}a_{+-}\big]+\sum_X 2Re\big[a^{P*}_{++}a_{++}+a^{P*}_{+-}a_{+-}\big], \label{f:f1Im}\\
  &g_{1L}+v_{1L} \sim Im(\mathcal{A}_{++,++}-\mathcal{A}_{+-,+-}) \sim \sum_X2\big[a^*_{++}a_{++}-a^*_{+-}a_{+-}\big]+\sum_X 2Re\big[a^{P*}_{++}a_{++}-a^{P*}_{+-}a_{+-}\big], \label{f:g1LIm}\\
  &h_{1T}+w_{1T} \sim Im\mathcal{A}_{++,--}\sim \sum_X \big[2a^*_{--}a_{++} + 2Re(a^{P*}_{--}a_{++})\big]. \label{f:h1TIm}
\end{align}
According to Eqs. (\ref{f:f1Im})-(\ref{f:h1TIm}), we can write down the following inequalities immediately, 
\begin{align}
  f_1(x)+u_1(x)\geq |g_{1L}(x)+v_{1L}(x)|. \label{f:fuinequality}
\end{align}



In the previous section, we did not introduce the definition of the parity-odd quark-quark correlator. Now we present the definition here.
In the previous discussions, we find it is convenient to separate the parity-odd quantities from the parity-even ones. One may argue that it is unnecessary to introduce the parity-odd correlator since correlator Eq. (\ref{f:correlatorE}) contains both the parity-even and parity-odd components. For us, we think it is necessary. This separation can not only help us to illustrate the properties of the parity-odd PDFs but also benefit the future studies, e.g., model calculations. In this case, we divide the correlator into two parts, one is the parity-even and the other is parity-odd.
\begin{align}
  \hat\Phi(k,p,S)\to \hat\Phi'(k,p,S)=\hat\Phi(k,p,S)+\hat\Phi^P(k,p,S), \label{f:Phitwo}
\end{align}
where $\hat\Phi(k,p,S)$ is defined in Eq. (\ref{f:correlatorE}). The parity-odd correlator can be defined as follows.


As shown in ref. \cite{Yang:2019gdr}, we introduce the parity-odd correlator according to the following arguments. First of all we know the vertex in weak theory is given by $\bar \psi\gamma^\mu(c_V^f\pm c_A^f\gamma^5)\psi$ where $c_V^f, c_A^f$ are the coupling constants. In the current theory, the vector current and the axial vector current are defined as $j^\mu=\bar \psi\gamma^\mu\psi, j^{\mu5}=\bar \psi\gamma^\mu\gamma^5\psi$. Based on these facts we can introduce the parity-odd correlator immediately by the replacement, $\psi \to \gamma^5 \psi$. 
This replacement indicates the parity-odd and parity-even correlators have similar forms and hence the parity-odd PDFs can be seen as the extensions of the parity-even ones. Since there is a factor $\gamma^5$ introduced in the correlator, the decomposition of the correlator with Dirac matrices must also have a $\gamma^5$. This consists with the previous discussion.

The operator definition of the parity-odd quark-quark correlator is given by
\begin{align}
  \hat\Phi^P(k,p,S)=\frac{1}{2\pi}\int d^4\xi e^{ik\xi}\Big(\langle N,S|[\bar\psi(0)\gamma^5]\ \psi(\xi)|N,S\rangle + \langle N,S|\bar\psi(0)\ [\gamma^5\psi(\xi)]|N,S\rangle \Big), \label{f:Phioddcorrelator}
\end{align}
We note here this definition may not be the unique definition of the parity-odd correlator. However, it seems that it is an economical one.
Under this circumstance, the one dimensional leading twist PDFs can be defined via
\begin{align}
  \hat\Phi^{P[\Gamma]}(x)&=\frac{1}{2\pi} \int d\xi^- e^{ixp^+\xi^-}
  \Big(\langle N,S|Tr\big[\bar\psi(0)\gamma^5 \Gamma \psi(0)\big]|N,S\rangle
  + \langle N,S|Tr\big[\bar\psi(0)\Gamma \gamma^5\psi(\xi^-)\big]|N,S\rangle \Big), \label{f:Phiodd1D}
\end{align}
where $\Gamma=\gamma^+, \gamma^+\gamma^5, i\sigma^{i+}\gamma^5$. We can insert into Eq. (\ref{f:Phiodd1D}) a complete set of intermediate states $\{|n\rangle\}$. Using the projector $P_+=\gamma^-\gamma^+/2$, we obtain $\bar\psi\gamma^+\psi=\sqrt{2}\psi^\dag_+\psi_+, \bar\psi\gamma^+\gamma^5\psi=\sqrt{2}\psi^\dag_+\gamma^5\psi_+$, where $\psi_+=P_+\psi$. Substituting these relations into Eqs. (\ref{f:Phiodd1D}) yields \cite{Jaffe:1983hp,Barone:2001sp}
\begin{align}
  \hat\Phi^{P[\gamma^+]}(x) =  \frac{1}{\sqrt{2}}\sum_n\delta\big((1-x)p^+-p_n\big)& \Big\{\big[|\langle p|P_R\psi_+(0)|n \rangle|^2-|\langle p|P_L\psi_+(0)|n \rangle|^2 \big] \nonumber \\
  -& \big[|\langle p|P_R\psi_+(0)|n \rangle|^2-|\langle p|P_L\psi_+(0)|n \rangle|^2\big]\Big\}, \label{f:u1prob}\\
  \hat\Phi^{P[\gamma^+\gamma^5]}(x) = \frac{1}{\sqrt{2}}\sum_n\delta\big((1-x)p^+-p_n\big)& \Big\{\big[|\langle p|P_R\psi_+(0)|n \rangle|^2+|\langle p|P_L\psi_+(0)|n \rangle|^2 \big] \nonumber \\
  -& \big[|\langle p|P_R\psi_+(0)|n \rangle|^2+|\langle p|P_L\psi_+(0)|n \rangle|^2 \big] \Big\}, \label{f:v1Lprob}
\end{align}
where $P_{R,L}=(1\pm\gamma^5)/2$. Clearly, the chiral-even one-dimensional parity-odd PDFs vanish. For a chiral-odd PDF, e.g., $h_{1T}$, it must company with another chiral-odd PDF or FF to contribute to the cross section. It can be shown chiral-odd PDFs do not have contributions from macroscopic view by applying the same method used in ref. \cite{Yang:2019gdr}. Disappearance of parity-odd PDFs can be seen as the equivalent effect of all the interactions in the parity-odd domains are averaged to zero. Quarks interact with fluctuation bubbles when go through the parity-odd domains. As a result, the parity-odd effects can be reduced completely if the quark interacts with all the fluctuation bubbles along the path.
In other words, these parity-odd PDFs which are induced by the fluctuations of the tunneling events are local quantities and vanish when sum over the intermediate states $|n\rangle$. This conclusion does not mean Eq. (\ref{f:phiP})-(\ref{f:phi5P}) are invalid. Assuming that interactions between interacting quarks and gauge fields in the parity-odd domains are not complete or the intermediate states $|n\rangle$ are not complete, these parity-odd PDFs does exist. In this case, parity-odd PDFs can be detected in specific processes.

To have the measurable effects of the parity-odd quantities, one needs to consider the non-averaged fluctuations of the tunneling events. Considering high energy reactions which PDFs participate in, one has the possibility to extract parity-odd PDFs when tunneling events emerge.
Though parity-odd PDFs and FFs are both induced by the tunneling events in parity-odd domains, they have some differences. Parity-odd FFs only survive on the event-by-event basis since the sum of all the hadrons in the final states can completely reduce the non-trivial $\theta$-vacuum tunneling effects by averaging all the fluctuation events. However, PDFs are limited in hadrons. The parity-odd effects are reduced completely from the averaging over all the interactions between the interacting quark and the fluctuation bubbles along the path in the hadron. As a result, the parity-odd FFs can be detected at moderate energy regions on the event-by-event basis while parity-odd PDFs are measurable at very high energy regions. Because relativistic time dilation can slow down the rate at which the quark interacts with the fluctuation bubbles in the parity-odd domains. In the following section we consider the jet production SIDIS process where asymmetries can be measured to study parity-odd PDFs.

\section{Spin asymmetries}\label{S:Applications}


To study the parity-odd PDFs, we consider the high energy SIDIS process in this section , $e^-(l)+N(p)\to e^-(l')+jet(k') +X(p_X)$. $l, l'$ are the momenta of incoming and outgoing electrons while $p, k'$ are the momenta of the target hadron and produced jet. $X (p_X)$ denotes undetected final states (momenta).
The differential cross section of the SIDIS can be written as the product of the leptonic tensor and the hadronic tensor,
\begin{align}
  d\sigma=\frac{\alpha_{em}^2e_q^2}{sQ^4}L^{\mu\nu}(l,l')W_{\mu\nu}(q,p,S,k')\frac{d^3l'd^3k'}{(2\pi)^32E_{l'}E_{k'}}, \label{f:crosssection}
\end{align}
where $\alpha_{em}$ is the fine structure constant, $e_q$ is the quark electric charge, $s=2l\cdot p$, $Q^2=-q^2$. A summation of flavors is  understood. The leptonic tensor is given by
\begin{align}
  L^{\mu\nu}(l,l')=2\left(l^\mu l^{\prime \nu}+ l^\nu l^{\prime \mu}-g^{\mu\nu} l\cdot l'\right)+2 i\lambda_e \varepsilon^{\mu\nu \alpha\beta}l_\alpha l'_\beta, \label{f:leptonictensor}
\end{align}
where $\lambda_e$ is the helicity of the electron. The hadronic tensor is defined as
\begin{align}
  W_{\mu\nu}(q,p,S,k')=\frac{1}{2\pi} \sum_X (2\pi)^4\delta^4(q+p-k'-p_X)\langle N,S|j_\mu(0)|k',X \rangle \langle k',X| j_\nu(0) |N,S\rangle, \label{f:hadronictensor}
\end{align}
where we do not distinguish the parity-odd and parity-even components and $j_\mu(0)$ denotes the electromagnetic current. Usually it is convenient to consider the integrated hadronic tensor $W_{\mu\nu}(q,p,S,k'_T)$ which is defined as
\begin{align}
  W_{\mu\nu}(q,p,S,k'_T)=\int \frac{dk'_z}{(2\pi)^32E_{k'}} W_{\mu\nu}(q,p,S,k'). \label{f:hadronicint}
\end{align}
Therefore the integrated differential cross section can be rewritten as
\begin{align}
  \frac{E_{l'}d\sigma}{d^3l'd^2k'_T}=\frac{\alpha_{em}^2e_q^2}{sQ^4}L^{\mu\nu}(l,l')W_{\mu\nu}(q,p,S,k'_T). \label{f:crosssectionint}
\end{align}

Since the higher twist contributions are suppressed by $1/Q$, in this paper we only consider the leading twist contributions. It has been shown, after collinear expansion, the hadronic tensor can be written as \cite{Liang:2006wp,Song:2010pf}
\begin{align}
  W_{\mu\nu}(q,p,S,k_T)=\frac{1}{2}Tr\left[\hat h_{\mu\nu} \hat\Phi'(x,k_T)\right], \label{f:hadronictrace}
\end{align}
where $\hat h_{\mu\nu}=\gamma_\mu\slashed n \gamma_\nu$, $\hat\Phi'(x,k_T)$ is the complete TMD correlator including both the parity-odd and parity-even correlators. Substituting the decompositions of these correlators Eqs. (\ref{f:phi})-(\ref{f:phioddP}) into Eq. (\ref{f:hadronictrace}), we can obtain the leading twist hadronic tensor,
\begin{align}
  W_{\mu\nu}(q,p,S,k_T) =& -2g_{T\mu\nu}\left(f_1(x,k_T)+\frac{\varepsilon_{T}^{kS}}{M}f_{1T}^\perp(x,k_T) + \lambda_h v_{1L}(x,k_T) + \frac{k_T\cdot S_T}{M} v^\perp_{1T}(x,k_T)\right) \nonumber\\
  &-2i\varepsilon_{T\mu\nu}\left(u_1(x,k_T)+\frac{\varepsilon_{T}^{kS}}{M}u_{1T}^\perp(x,k_T) + \lambda_h g_{1L}(x,k_T) + \frac{k_T\cdot S_T}{M} g^\perp_{1T}(x,k_T)\right),\label{f:hadronicresult}
\end{align}
where $g_{T\mu\nu}=g_{\mu\nu}-\bar n_\mu n_\nu -\bar n_\nu n_\mu$, $\varepsilon_{T\mu\nu}=\varepsilon_{\alpha\beta\mu\nu}\bar n^\alpha n^\beta$. It is can be easily checked that the hadronic tensor satisfies the current conservation $q^\mu W_{\mu\nu}=q^\nu W_{\mu\nu}=0$.

By carrying out the contraction of the leptonic tensor and the hadronic tensor, in the $\gamma^*N$ frame the differential cross section can be written as
\begin{align}
 \frac{E_{l'}d\sigma}{d^3l'd^2k'_T}=\frac{4\alpha_{em}^2}{sQ^4y^2}
 &\Big[A(y)f_1 - C(y)\lambda_e u_1 + A(y)\lambda_h v_{1L} -C(y)\lambda_e \lambda_h g_{1L} \nonumber\\
 &-|S_T|k'_{TM}\Big(A(y)f_{1T}^\perp - C(y)\lambda_e u^\perp_{1T} \Big) \sin(\phi-\phi_S) \nonumber\\
 &-|S_T|k'_{TM}\Big(A(y)v_{1T}^\perp - C(y)\lambda_e g^\perp_{1T} \Big) \cos(\phi-\phi_S)
 \Big], \label{f:crosssectionresult}
\end{align}
where $A(y)=1+(1-y)^2, C(y)=2y-y^2, y=p\cdot q/p\cdot l, k'_{TM}=|\vec k'_T|/M$. $\phi, \phi_S$ are azimuthal angles of the jet and hadron spin with respect to the lepton-hadron plane, respectively. We can see that the SIDIS differential cross section is modified by the parity-odd PDFs.
From Eq. (\ref{f:crosssectionresult}) we can calculate the following spin asymmetries.

\subsection{Single spin asymmetry}

We first consider the single spin asymmetries. We assume that the target hadron is polarized while the electron is unpolarized.
For longitudinal polarized hadron, the single spin asymmetry is defined as,
\begin{align}
  A_{L} &= \frac{d\sigma(\lambda_e=0,+)-d\sigma(\lambda_e=0,-)}{2d\sigma_{unp}}, \label{f:ALdef}
\end{align}
where subscript $L$ denotes longitudinal polarized hadron, $+, -$ denote the positive and negative helicities of the hadron, $d\sigma_{unp}$ denotes the unpolarized cross section, $\frac{E_{l'}d\sigma_{unp}}{d^3l'd^2k'_T}=\frac{4\alpha_{em}^2}{sQ^4y^2}A(y)f_1(x, k_T)$. Substituting the differential cross section Eq. (\ref{f:crosssectionresult}) into this definition, we can obtain,
\begin{align}
  A_{L} = \frac{v_{1L}(x,k_T)}{f_1(x,k_T)}. \label{f:ALres}
\end{align}
We can see that longitudinal polarization of the hadron is determined by the parity-odd PDF $v_{1L}$. This kind of asymmetry can be used to extract $v_{1L}$. Hereafter, one should know there are summations of flavors in these asymmetries.

For the transverse polarized target, the SSA is defined as,
\begin{align}
  A_{T}^{U,L} &= \frac{d\sigma(\lambda_e=0,\uparrow^{U,L})-d\sigma(\lambda_e=0,\downarrow^{U,L})}{2d\sigma_{unp}}. \label{f:ATdef}
\end{align}
where subscript $T$ denote transversely polarized hadron, superscripts $U, L$ denote the unpolarized and longitudinal polarized quark distributions in a transversely polarized hadron. For the unpolarized distribution of quarks in a transversely polarized hadron, we have
\begin{align}
  A^U_{T}=-k'_{TM}\frac{f_{1T}^\perp(x,k_T) }{f_1(x,k_T)}\sin(\phi-\phi_S), \label{f:sin}
\end{align}
where superscript $U$ denotes unpolarized case. We see Eq. (\ref{f:sin}) corresponds to the famous Sivers effect because it results from the Sivers function $f^\perp_{1T}$. The Sivers function is interpreted as the distribution of a unpolarized quark  distribution inside a transversely polarized hadron.
We see that $A^U_{T}$ is corresponding to the azimuthal asymmetry $\langle \sin(\phi-\phi_S)\rangle$ which can be measured in experiments to extract the corresponding PDF, $f^\perp_{1T}$.
For the longitudinal polarization distribution of quarks in a transversely polarized hadron, we have
\begin{align}
  A^L_{T}=-k'_{TM}\frac{v_{1T}^\perp (x,k_T)}{f_1(x,k_T)}\cos(\phi-\phi_S), \label{f:cos}
\end{align}
where superscript $L$ denotes longitudinal polarized case. We see $A^L_{T}$ obtains contribution from the parity-odd PDF $v_{1T}^\perp$. It can be seen $A^L_{T}$ corresponds to the azimuthal asymmetry $\langle \cos(\phi-\phi_S) \rangle$ which can be used to extract the corresponding PDF, $v^\perp_{1T}$.

From Eq. (\ref{f:crosssectionresult}) we can see that the SIDIS cross section can be reduced to the DIS one by integrating over $k_T$. Hence only the one dimensional PDFs have contributions to the cross section. It is interesting to see that parity-odd PDF $u_1$ even can be measured when the target is unpolarized. We define
\begin{align}
  A_{\lambda_e} &= \frac{d\sigma(\lambda_e=+1,0)-d\sigma(\lambda_e=-1,0)}{2d\sigma_{unp}}. \label{f:AULdef}
\end{align}
Therefore,
\begin{align}
  A_{\lambda_e}=-\frac{C(y)}{A(y)}\frac{u_1(x)}{f_1(x)}. \label{f:ALU}
\end{align}
$A_{\lambda_e}$ can be measured with the polarized beam in high energy (SI)DIS. We can see that Eq. (\ref{f:ALU}) is similar to the asymmetry measured in parity violating DIS (PVDIS) \cite{Cahn:1977uu},
\begin{align}
  A^{PV}&=\frac{ C_{1q}+R_yC_{2q}}{e_q^2 }\frac{f_1(x)}{f_1(x)} ,\label{f:APVNin}
\end{align}
where $C_{iq}, i=1,2$, are coupling constants, $R_y=C(y)/A(y)$. A summation of flavors is  understood. In PVDIS, the parity violation is global because it comes from weak interaction. Hence the parity-odd effects are embodied in coupling constants when $f_1$ is parity-even. For $A_{\lambda_e}$, the parity-odd effects come from the parity-odd PDF $u_1$ rather than the coupling constants.

\subsection{Double spin asymmetry}

Since both the hadron spin and the electron spin are involved in the cross section, Eq. (\ref{f:crosssectionresult}), we consider the double spin asymmetries. For the longitudinal polarized hadron case, we define the double spin asymmetry as \cite{Kanazawa:2014tda}
\begin{align}
  A_{LL}= \frac{\big[d\sigma(\lambda_e=+1,+)-d\sigma(\lambda_e=-1,+)\big] -\big[d\sigma(\lambda_e=+1,-)-d\sigma(\lambda_e=-1,-)\big]}{4d\sigma_{unp}}. \label{f:doubleLLdef}
\end{align}
Substituting the differential cross section Eq. (\ref{f:crosssectionresult}) into this definition, we can obtain,
\begin{align}
  A_{LL}=-\frac{C(y)}{A(y)}\frac{g_{1L}(x,k_T)}{f_1(x,k_T)}. \label{f:doubleLL}
\end{align}
The PDF $g_{1L}$ denotes the longitudinal polarized quark distribution in a longitudinal polarized. It is strongly related the hadron spin with the relation $g_1 = \frac{1}{2}\sum_qe_q^2 g^q_{1L}$, $g_1$ is the structure function.  For the transversely polarized hadron case, we define the double spin asymmetry as
\begin{align}
  A_{LT}^{U,L}= \frac{\big[d\sigma(\lambda_e=+1,\uparrow^{U,L})-d\sigma(\lambda_e=-1,\uparrow^{U,L})\big] -\big[d\sigma(\lambda_e=+1,\downarrow^{U,L})-d\sigma(\lambda_e=-1,\downarrow^{U,L})\big]}{4d\sigma_{unp}}. \label{f:doubleLTdef}
\end{align}
Substituting the differential cross section Eq. (\ref{f:crosssectionresult}) into this definition, we can obtain,
\begin{align}
  &A_{LT}^{U}= k'_{TM}\frac{C(y)}{A(y)}\frac{u_{1T}^\perp(x,k_T)}{f_1(x,k_T)}\sin(\phi-\phi_S), \label{f:doubleLTU} \\
  &A_{LT}^{L}= k'_{TM}\frac{C(y)}{A(y)}\frac{g_{1T}^\perp(x,k_T)}{f_1(x,k_T)}\cos(\phi-\phi_S). \label{f:doubleLTL}
\end{align}
From Eqs.(\ref{f:ALres}), (\ref{f:cos}), (\ref{f:ALU}) and (\ref{f:doubleLTU}), we see that these spin asymmetries are all generated by the parity-odd PDFs. They corresponds to $v_{1L}$, $v_{1T}^\perp$, $u_{1}$ and $u_{1T}^\perp$, respectively. They provide us a effective way to measure the parity-odd PDFs. Since we only consider the jet production SIDIS process, only chiral-even PDFs are taken into account.

\section{summary}\label{S:summary}

The true physical QCD vacuum state is a linear superposition of the $n$ vacua states with different topological numbers. The vacuum to vacuum transition amplitude is determined by the topological charge (density) which is equal to the $\theta$-term. The $\theta$-term in the QCD Lagrangian is parity violated.
Because of the configuration of the gauge fields, the tunneling events can induce the local parity-odd domains. Those interactions that occur in these domains can be affected by the tunneling events. In this paper, we introduce the parity-odd PDFs in order to describe the parity-odd structures inside the hadron. Among these 16 PDFs obtained by decomposing the quark-quark correlator, 8 are parity-even and the other are parity-odd PDFs for spin-1/2 hadrons. They have a one-to-one correspondence.
In this paper we calculate the SIDIS process to give the measurable quantities of parity violation. From Eqs.(\ref{f:ALres}), (\ref{f:cos}), (\ref{f:ALU}) and (\ref{f:doubleLTU}), we see that these spin asymmetries are all generated by the parity-odd PDFs. They provide us a effective way to measure the parity-odd PDFs.

To summarize this paper, we propose a hypothesis that the small nEDM is due to the averaging over the tunneling events giving rise to the electric charge separation. As argued in ref. \cite{Faccioli:2004ys}, during the tunneling process, the $\theta$-term can generate an effective repulsion between matter and antimatter. Supposing each of these tunneling events can form a EDM $\vec d_i$, we can see that the averaging over all of the EDMs can be zero $\sum_i \vec d_i=0$ because the directions of these EDMs are completely random without external fields. In other words, intermediate states $|p\pi\rangle$ can induce the EDM between pion and proton in each parity-odd domain in neutron. However, these processes are also random without external fields and does vanish when sum over all of them.


\end{document}